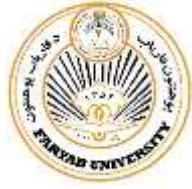

# Ineffectiveness of Alien Terms Interference in a Culture of Multilingual Counties

**Mohammad Ibrahim Qani**
**Senior Assistant Professor (Pohanmal)**
**Mobile:** +93794555300

**E-mail: ibrahim.qani@gmail.com & ibrahim.qani2021@faryab.edu.af**

**2025**

**Abstract**
Language serves as a foundation of cultural identity, deeply entangled with the social and historical contexts of a community. This paper examines the ineffectiveness of interference by alien words within a culture. Drawing on sociolinguistic theories and case studies from diverse linguistic environments, it is argued that the forced introduction or adoption of foreign lexicon often fails to achieve its intended socio-cultural objectives. Instead, indigenous languages demonstrate resilience, adapting to or resisting external influences through unique strategies. The effectiveness of this research highlights the futility of attempting to impose linguistic uniformity and underscores the importance of understanding local cultural dynamics in preserving linguistic heritage. This pure language understanding directly relates to translation knowledge where linguists and translators need to work and research to eradicate misunderstanding. Misunderstandings mostly appear in non-equivalent words because there are different local and internal words like food, garment, cultural and traditional words, and others in every notion. Truly, most of these words do not have an equivalent in the target language and these words need to be worked and find their equivalent in the target language to fully understand both languages. The purpose of this research is to introduce the challenges and ineffectiveness of cultural influences in different notions where people do not see the facts of cultural enrichment. However, some of these ineffectiveness have been clearly mentioned in this research but some effective ways have also been dictated.

**Methods of Research**

In this article, I used the APA style and the usual old research method which is called Library Research (data collection), online libraries, PDF books, available research papers in reliable journals and websites, and some other sources. This article covers the analysis of rendering some sort of technical and scientific words from the source language into the target language.

**Introduction**

Language serves as a vital conduit of culture, encapsulating the nuances and values inherent to a particular society. However, the introduction of alien words into a language often sparks debates regarding the potential dilution or transformation of cultural identity. This essay contends that such interference is largely ineffectual, as historical evidence reveals that cultures exhibit remarkable resilience and adaptability in the face of lexical changes. The infusion of foreign terms often mirrors the dynamic nature of language itself, where borrowing is a source of enrichment rather than cultural erosion. While some may argue that the adoption of alien words undermines authentic expression within a language, a closer examination reveals that cultures are not monolithic entities; they thrive through synthesis and innovation. Thus, understanding the interplay between language and culture provides insight into the broader societal implications of linguistic interference, ultimately showcasing the enduring strength of cultural identities.

**Definition of alien words and their role in cultural exchange**

The incorporation of alien words within a language signifies not only linguistic exchange but also a deeper cultural interplay, often revealing the complexities and tensions in such interactions. Alien words, which are terms adopted from one language into another, reflect a societys adaptive response to external influences, and they often carry cultural significance and connotations that extend beyond their literal meanings. This phenomenon underscores a dual process: while alien words can introduce innovative concepts and frameworks, they may also create discord within the receiving culture, diluting indigenous linguistic identities and traditions. Critics argue that this interference ultimately undermines cultural integrity. In examining the role of alien words in cultural exchange, Pellizzioni (2016), highlights the neglect of the intricate balance between adaptation and dominance, further reinforcing the argument that reliance on external lexicons often complicates cultural cohesion (p. 15). Nevertheless, as notes, the negotiation of such influences can lead to hybrid forms that challenge traditional norms and create new avenues for cultural evolution (Nordberg, 2014, p. 124).

**Historical Context of Language Interference**

The historical context of language interference provides a significant backdrop for understanding the ineffectiveness of alien words within a culture. Language, shaped by cultural evolution, often experiences interference due to contact with other languages, leading to the incorporation of foreign terms. However, the philosophical frameworks underlying this process can illuminate the complexities involved. For instance, the concept of mnidoo-worlding in Ojibwe Anishinaabe philosophy emphasizes an interconnectedness of beings, where external influences are perceived through a relational lens (Manning, 2017, pp. 12, 18). This perspective challenges the simplistic view of language acquisition, suggesting that cultural entities resist co-option by foreign words that do not align with their intrinsic values. Additionally, the study of phonetic symbolism indicates that people may inherently associate certain sounds with meanings, reinforcing the idea that alien words struggle to find resonance within established cultural paradigms. As such, historical interactions shape not only the languages themselves but also the cultural milieus receptivity to foreign elements (Tarr, 1979, p.87).

> A. Examples of alien words in various cultures and their initial impact
>
> The infusion of alien words into various cultures often initiates a series of complex social and linguistic dynamics, revealing the nuances of cultural interaction and transformation. For instance, when English words like pizza and computer entered non-English-speaking contexts, they brought with them not only new concepts but also shifts in culinary practices and technological understanding. Initially, these imports can enhance communication and contemporary relevance; however, they may inadvertently challenge traditional lexicons, leading to a dilution of cultural heritage. Moreover, the reliance on foreign terminology can create barriers to understanding, particularly among older generations who may resist adopting unfamiliar terms. As articulated in the context of invasive alien plants, similar patterns of disruption can emerge, complicating the management of cultural integrity amidst external influences. Thus, while alien words may serve immediate practical purposes, their long-term consequences often reflect a deeper cultural unease (Masunungure, 2020, p. 63).

Linguistic interference refers to the influence of one language on another, which often occurs in situations of language contact, such as bilingualism, multilingualism, or language acquisition. This phenomenon can manifest at various linguistic levels, including phonetic, lexical-semantic, morphological, syntactic, and stylistic1. However, the forced introduction or adoption of alien words, i.e., foreign lexicon, can be largely ineffective for several reasons:

Cultural Resilience: Cultures often show a strong resistance to external linguistic influences, preserving their own linguistic heritage and identity.

Adaptation and Localization: Indigenous languages tend to adapt foreign words to fit phonological and morphological patterns, diminishing the 'alien' aspects of the words.

Sociolinguistic Factors: The acceptance of foreign words is influenced by various sociolinguistic factors such as prestige, necessity, and social acceptance.

Historical and contemporary case studies illustrate how communities have either resisted or adapted to such linguistic interference based on their unique cultural contexts.

For a more detailed understanding, you can refer to this scholarly article on linguistic interference, which explores the different types and their peculiarities. Also, exploring the essay on language interference provides insights into the challenges and outcomes of such interference (Kamolovna, 2023, 191).

**Linguistic Resistance and Adaptation**

In examining the phenomenon of linguistic resistance and adaptation, it is essential to recognize how cultures navigate the influx of alien words while preserving their intrinsic identity. Despite the pressures of globalization that often lead to superficial assimilation of foreign terms, communities demonstrate a remarkable capacity for adaptation, effectively reshaping these interlopers into forms that resonate with their linguistic frameworks. This process can be likened to mnidoo-worlding, where new ideas and expressions are interwoven with existing cultural narratives, reflecting a co-responsiveness among diverse elements of existence (Manning, 2017, p. 101). Such linguistic practices illustrate not only the resilience of cultural identity but also facilitate a nuanced dialogue between legacy and innovation. Ultimately, this dynamic underscores that the effectiveness of alien word interference is considerably diminished by the robust mechanisms of adaptation, suggesting that cultural integrity emerges through transformation rather than outright acceptance (McCleary, 2019, p. 67).

  A. Mechanisms through which cultures resist or adapt to alien words

Understanding the mechanisms through which cultures resist or adapt to alien words involves recognizing the inherent resilience and flexibility of linguistic frameworks. Cultures often employ various strategies to maintain their linguistic integrity while incorporating foreign terms, demonstrating a dual capacity for adaptation and resistance. For example, languages may create neologisms or linguistic hybrids that reflect existing phonetic and grammatical norms, thereby preserving cultural identity while accommodating external influences. Furthermore, as seen in the analysis of mahr agreements and religious divorce across multiple jurisdictions, systems tend to resist direct and command-oriented interventions, leading to outcomes that can be ineffective or incoherent (Cummings, 2023, p. 165). This theme resonates in modern contexts as well; just as military strategies need to adapt to local realities, cultural interactions require a nuanced understanding of how alien words can either enhance or undermine indigenous languages. Such dynamics illustrate the complex interplay of acceptance and rejection within cultural linguistic evolution (Metz, 2004, p 95).

## Ineffectiveness of Interference of Alien Words in a Culture

The phenomenon of alien words, often termed loanwords, involves the adoption of terms from one language into another. While such linguistic borrowing can enrich a language, the interference of alien words may also lead to issues in cultural identity and communication. This paper will explore the ineffectiveness of alien words when integrated into a culture, focusing on how they can disrupt native linguistic structures and contribute to confusion rather than clarity.

## Cultural Identity

Language is a critical component of cultural identity. As researchers like Edward Sapir have noted, language shapes thought and influences social structures (Sapir, 1921, p. 54). The introduction of alien words can dilute the richness of a native language, leading to a loss of cultural specificity. The encroachment of foreign terms can foster a feeling of disconnection among speakers, especially when they are unable to find equivalent expressions in their native tongues (Crystal, 1987, p. 78). This suggests that, rather than enhancing communication, alien words can create barriers that fragment cultural understanding.

## Communication Clarity

The introduction of alien words often presents challenges in communication. As observed by Akhmanova (1966), alien terminology may confuse speakers who are not familiar with the foreign concepts being introduced, leading to misunderstandings. For example, in the context of modernization and globalization, certain technical words may be adopted without clear definitions within the cultural milieu. This can result in ineffective communication and a lack of comprehension among the local populace (Haugen, 1950, p. 178).

## Linguistic Integrity

Linguistic integrity is another aspect affected by the presence of alien words. Excessive reliance on loanwords can undermine the grammatical structure and phonetic patterns of a language. According to Mufwene (2001), languages are dynamic systems that evolve over time. However, this evolution should respect the intrinsic characteristics of the original language. When alien words dominate, they may contribute to a hybridization that neglects the foundational elements of the culture's language.

The ineffectiveness of alien words in a culture underscores the delicate balance between linguistic evolution and cultural preservation. While some integration of foreign terms is inevitable and can be beneficial, excessive interference can lead to cultural fragmentation and communication barriers. A critical approach to the adoption of alien words is essential to ensure that the integrity and richness of a culture's language are maintained.

**Interference in the Culture of Multilingual Countries**

The phenomenon of alien terms in multilingual countries brings a unique set of challenges and complexities. While the borrowing of words from one language to another can enhance communicative exchanges, it can also lead to cultural dissonance and misunderstandings, particularly in multilingual contexts. This paper explores the ineffectiveness of alien term interference in the cultures of multilingual countries, emphasizing its impact on identity, communication, and linguistic integrity.

**Cultural Identity**

In multilingual societies, language plays an essential role in expressing and maintaining cultural identity. Alien terms—words borrowed from one language into another—can disrupt this identity by overshadowing indigenous languages. According to Heller (2007), the use of foreign terms may create a perception of elitism or exclusion, making speakers of indigenous languages feel marginalized (p. 65). This alienation can hinder the preservation of cultural heritage, as individuals may prioritize languages associated with perceived social or economic advantages over their native tongues.

**Communication Clarity**

Communication in multilingual contexts is inherently complex, and the introduction of alien terms can exacerbate misunderstandings. As noted by Auer (2013), the mixing of languages can lead to semantic ambiguity, making it difficult for speakers to convey precise meanings (p. 112). This can create confusion, particularly when alien terms lack direct equivalents in the local languages. For example, in countries like India, where English permeates various domains, the excessive use of English loanwords can interfere with effective communication among speakers of regional languages, thus inhibiting mutual understanding (Bhatt, 2010, p. 455).

**Linguistic Integrity**

The impact of alien terms on linguistic integrity in multilingual countries is a significant concern. Mufwene (2001) posits that the integration of external vocabulary often dilutes the structural uniqueness of a language (p. 16). In multilingual settings, where code-switching occurs frequently, the prevalence of alien terms may lead to hybrid vernaculars that deviate from traditional linguistic norms. This hybridization can compromise the richness of distinct languages and dialects, causing a disconnection from cultural roots (Heller, 2007, p. 68).

The ineffectiveness of alien terms in the cultural landscape of multilingual countries underscores the need for careful consideration in linguistic borrowing. While alien terms can facilitate communication, their interference often leads to cultural fragmentation, confusion, and erosion of linguistic integrity. Therefore, fostering an environment that respects and revitalizes native languages is essential for maintaining cultural identity and enhancing communication in multilingual settings.

**Conclusion**

In conclusion, the ineffectiveness of interference from alien words within a culture reflects deeper dynamics concerning identity and autonomy. Cultures maintain resilience against external linguistic incursions, suggesting that such influences often serve more to highlight cultural distinctions than to integrate genuinely. The implications of this resistance can be observed in various domains, especially as societies face increasing globalization pressures. The capacity of a culture to absorb or reject a foreign lexicon is indicative of its strength and coherence. Furthermore, the cultural construction of identity, as critiqued in the discourse surrounding new materialism, underscores a conflict between embracing fluidity and maintaining foundational values (Boggio, 2005, p. 18). This duality is significant, as the tensions of alien terminology can ultimately compromise local expressions and meanings, thereby exacerbating existing inequalities and inhibiting genuine dialogue. Overall, understanding these aspects reinforces the necessity for nurturing robust cultural frameworks in an increasingly interconnected world (Pellizzioni, 2016, p. 42). Summary of the ineffectiveness of alien words in altering core cultural identities, he attempts to alter core cultural identities through the introduction of alien words often proves ineffective, as language serves not merely as a collection of terms but as a vessel for the values, beliefs, and practices intrinsic to the culture. Alien words, while they may momentarily pique interest or trend within certain social circles, seldom take root in the foundational aspect of a community's identity. For example, borrowed terminology from foreign languages may be adopted in specific contexts such as fashion or technology but fails to transform the deeper cultural narratives that define a group's essence. Consequently, resistance to these external influences emerges, reinforcing a sense of cultural pride and belonging among individuals. Therefore, while the surface of a language may change, the underlying cultural frameworks remain steadfast, ultimately illustrating the resilience of core identities against the superficial impact of alien linguistic elements.